\documentclass{PoS}
\usepackage{float}
\usepackage{subcaption}

\title{Creating a high-resolution picture of Cygnus with the Cherenkov Telescope Array}

\ShortTitle{Cygnus with CTA}

\author{\speaker{Amanda Weinstein}$^{a}$,  {Ester Aliu}$^{b}$, {Sabrina Casanova}$^{c}$, {Tristano Di Girolamo}$^{d}$, {Michael Dyrda}$^{e}$, {Joachim Hahn}$^{c}$, {Pratik Majumdar}$^{f}$, {Jerome Rodriguez}$^{g}$, {Luigi Tibaldo}$^{h}$,
  for the CTA Consortium\footnote{Full consortium author list at http://cta-observatory.org}\\
        E-mail: \email{amandajw@iastate.edu}

{\footnotesize
$^{a}$ Department of Physics and Astronomy, Iowa State University, USA;
$^{b}$ Departament d'Astronomia i Meteorologia, Institut de Ci{\`{e}}cies del Cosmos, Universitat de Barcelona, IEEC-UB, Mart\'{i} i Franqu\`{e}s 1, E-08028 Barcelona, Spain;
;
$^{c}$ Max-Planck-Institut f\"{u}r Kernphysik, Heidelberg, Germany;
$^{d}$ INFN Sezione di Napoli and Dipartimento di Scienze Fisiche, Universit\`{a} degli Studi di Napoli "Federico II", Italy;
$^{e}$ Institute of Nuclear Physics, Polish Academy of Sciences, Poland;
$^{f}$ Saha Institute of Nuclear Physics, India;
$^{g}$ IRFU/SAp, CEA Saclay, France;
$^{h}$ Kavli Institute for Particle Astrophysics and Cosmology, Department of Physics and SLAC National Accelerator Laboratory, Stanford University, USA}
	
}	

\abstract{The Cygnus region hosts one of the most remarkable star-forming regions in the Milky Way. Indeed, the total mass in molecular gas of the Cygnus X complex exceeds 10 times the total mass of all other nearby star-forming regions. Surveys at all wavelengths, from radio to \g-rays, reveal that Cygnus contains such a wealth and variety of sources---supernova remnants (SNRs), pulsars, pulsar wind nebulae (PWNe), H\,{\tiny II} regions, Wolf-Rayet binaries, OB associations, microquasars, dense molecular clouds and superbubbles---as to practically be a galaxy in microcosm. The \g-ray observations along reveal a wealth of intriguing sources at energies between 1 GeV and tens of TeV. However, a complete understanding of the physical phenomena producing this \g-ray emission first requires us to disentangle overlapping sources and reconcile discordant pictures at different energies. This task is made more challenging by the limited angular resolution of instruments such as the Fermi Large Area Telescope, ARGO-YBJ, and HAWC and the limited sensitivity and field of view of current imaging atmospheric Cherenkov telescopes (IACTs). The Cherenkov Telescope Array (CTA), with its improved angular resolution, large field of view, and order of magnitude gain in sensitivity over current IACTs, has the potential to finally create a coherent and well-resolved picture of the Cygnus region between a few tens of GeV and a hundred TeV. We describe a proposed strategy to study the Cygnus region using CTA data, which combines a survey of the whole region at $65^{\circ} < l < 85^{\circ}$ and $-3.5^{\circ} < b < 3.5^{\circ} $ with deeper observations of two sub-regions that host rich groups of known \g-ray sources.}

\FullConference{The 34th International Cosmic Ray Conference,\\
		30 July- 6 August, 2015\\
		The Hague, The Netherlands}

\def\g{$\gamma$}
\newcommand{\F}{\textit{Fermi} LAT}

\begin{document}

\section{Introduction}

Among the star-forming regions of the Milky Way, the Cygnus region stands out as a superb laboratory for cosmic ray production, acceleration and transport.  Surveys of the region at all wavelengths, from radio to \g-rays, reveal a galaxy in microcosm, filled with supernova remnants (SNRs), pulsars, pulsar wind nebulae (PWNe), H\,{\tiny II} regions, Wolf-Rayet binaries, many OB associations, microquasars, dense molecular clouds and superbubbles, and even a \g-ray source that testifies to the presence of cosmic ray sources within superbubbles \cite{2011Sci...334.1103A}.
This last lies in the nearby ($d \approx 1.5$ kpc) Cygnus X complex, ten times as rich in molecular gas as all the other nearby star-forming regions  \cite{2006A&A...458..855S,2007A&A...474..873S}.
Multi-messenger data from IceCube and AMANDA data hints at a neutrino signal from the Cygnus region, which, if confirmed, would be the unambiguous signature of powerful
proton acceleration above tens of TeV \cite{2013APh....43..155H}.

Cygnus also boasts diffuse VHE \g-ray emission in excess of that predicted by modeling the interactions of the large-scale
cosmic-ray population with interstellar matter and radiation fields in the Galaxy \cite{2011Sci...334.1103A,2012A&A...538A..71A}.
This diffuse excess may indicate either a local overabundance of freshly-injected cosmic rays or an unresolved population of faint  very high energy (VHE; $E > 100$ GeV) \g-ray-bright sources.
Finally, the Cygnus region hosts two of the most remarkable high-mass X-ray
binaries known: Cyg X-3 and Cyg X-1.
Very sensitive observations of the Cygnus binaries in \g~rays could permit us to detect and probe the origin of VHE emission in microquasars and thus access the physics of the jets and
their interaction with the environment.
\cite{2007MNRAS.382..367B,2009A&A...507..241P}.


\subsection{Current Status}

In Cygnus, the brightest portion of the Northern \g-ray sky, we find a wealth of intriguing \g-ray sources at energies between 1 GeV and tens of TeV.
However, the picture provided by the various \g-ray observatories---\g-ray satellites such the Fermi Large Area Telescope (\F~) and AGILE at GeV energies, imaging atmospheric Cherenkov telescopes (IACTs) such as HEGRA, MAGIC and VERITAS at hundreds of GeV, and Milagro and ARGO-YBJ \cite{2012ApJ...745L..22B}  at TeV energies---is complex and at times discordant.
Beginning at 1 GeV \F~ detects an extended excess of hard \g-ray emission, the so-called Fermi cocoon \cite{2011Sci...334.1103A}. ARGO-YBJ and Milagro have detected sources that they associate with the cocoon at higher energies\cite{2014ApJ...790..152B}.
AGILE also reports a tentative detection of the X-ray persistent
stellar mass black-hole binary Cyg X-1 in a flaring state \cite{2013ApJ...766...83S}.

Surveys and targeted snapshots by IACTs present a different view.  In addition to MAGIC's reported detection of Cyg X-1 during a flare, they also identify a set of more compact extended sources:
TeV J2032+4130, the first unidentified and extended source ever discovered at very high energies, VER J2019+407, likely linked to radio-, X-ray-, and \g-ray-bright SNR $G78.2+2.1$, and VER J2019+37, linked to MGRO J2019+37 and one of the hardest-spectrum sources ever detected \cite{2002A&A...393L..37A,2013ApJ...770...93A,2007ApJ...665L..51A,
2014ApJ...788...78A}. Only upper limits could be set for the other well-known microquasar in the region, Cyg X-3, which is located in a much denser environment that Cyg X-1 \cite{2010ApJ...721..843A,2013ApJ...779..150A}.
Conspicuous by its absence is any IACT detection of the Cygnus cocoon.


We must disentangle these overlapping \g-ray sources if we are to definitely pinpoint the physical phenomena behind them.
This task is made more challenging by the limited angular resolution of instruments such as the \F~, ARGO-YBJ, and HAWC and the limited sensitivity and field of view of both past and current IACTs.
Establishing whether freshly-accelerated cosmic rays or unresolved faint sources \cite{2008APh....29...63C,2013APh....43..317D} produce the bright VHE \g-ray diffuse emission detected by MILAGRO \cite{2007ApJ...658L..33A,2008ApJ...688.1078A} also requires more sensitive instruments with finer angular resolution.

\section{Role of CTA}

CTA has the capabilities needed for a fruitful study of Cygnus \cite{2013APh....43....1H}.  A factor of ten improved sensitivity over current IACTs provides the statistics needed for detailed morphological and spectral studies as well as studies of transient phenomena from high-mass X-ray binaries. The large ($7-8^{\circ}$) field of view (FoV) is well-suited to surveys and boosts sensitivity to extended and diffuse sources.  Finally,  CTA's angular resolution far surpasses that of current \g-ray observatories, including HAWC, between a few hundred GeV and $\sim 100$ TeV, which will help separate out overlapping sources and pinpoint low-energy counterparts.
CTA's planned program of Key Science Programs, or KSPs, provides for a survey of the region $65^{\circ} < l < 85^{\circ}$ and $-3.5^{\circ} < b < 3.5^{\circ} $ down to 2.7 mCrab under the aegis of the Galactic Plane Survey KSP.  The Star-forming Regions KSP will augment this survey with deep observations of two sub-regions that host the richest and most interesting groups of known $\gamma$-ray sources: the region surrounding Cygnus OB2/Fermi cocoon and that surrounding Cygnus OB1/MGRO J2019+37.
We discuss the projected impact of these observations below.

\subsection{Cygnus OB2 and the Fermi Cocoon}\label{sec:cygob2}

Source confusion presents a major challenge when trying to pinpoint the source of the energetic particles flooding Cygnus-X.
The favored option is the collective action of star-wind shocks in the stellar clusters, but contributions from SNR~G78.2$+$2.1 or a PWNe cannot be completely ruled out.
Likewise, the energy spectra of the \F~cocoon, ARGO~J2031$+$4157, and MGRO~J2031+41 (hard \g-~ray excesses of roughly $\sim 2^\circ$ extension co-located between Cyg~OB2 and SNR~G78.2+2.1) connect relatively smoothly, bolstering the interpretation that these are detections of the same source at three different energies. Yet the coarser angular resolution of ARGO-YBJ and Milagro means that these instruments must rely on IACT measurements to correct the VHE spectrum for contributions both VER J2019+407 and TeV J2032+4130, which lies near Cyg OB2 and surrounds a \g-~ray pulsar that may itself be part of a binary system \cite{2015arXiv150201465L}.
This incurs an inflated systematic uncertainty due to cross-calibration issues \cite{2014ApJ...790..152B} and is further exacerbated by the non-detection of the cocoon by current IACTs, likely to due background-subtraction effects connected to the small field of view.
A more self-consistent treatment of the problem requires an instrument, like CTA, that can both detect and resolve all sources in the region.

\begin{figure}[htbp!]
\centering
\includegraphics[width=0.9\textwidth]{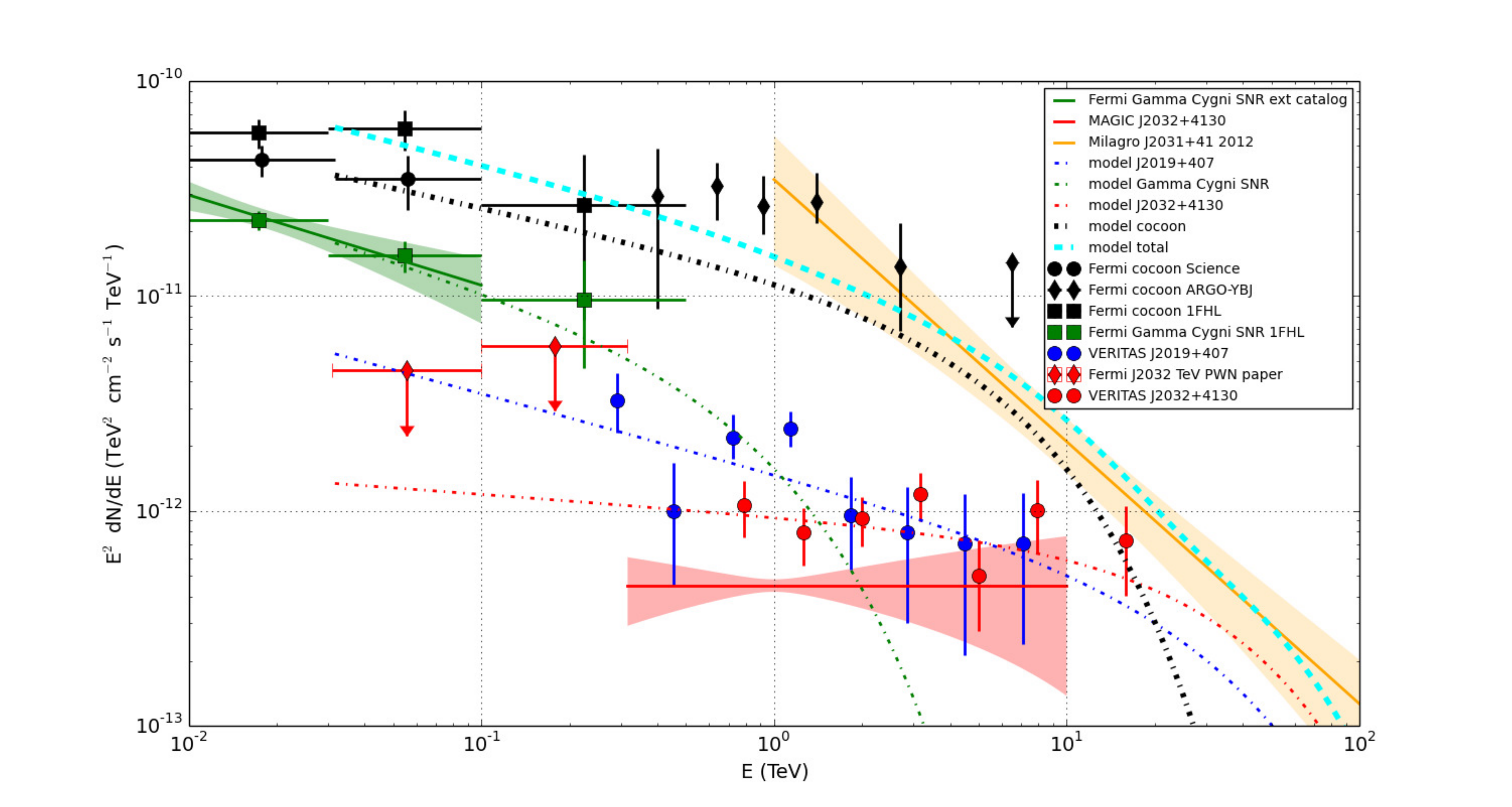}
\caption{\small Best-measured spectra above 10 GeV for all known sources in this region apart from the microquasar Cyg X-3 and the contributions from \g-ray pulsars, whose low energy cutoffs make them insignificant at these energies.  Dashed lines show plausible extrapolations of the fitted spectra.  Two independent measurements by VERITAS (red points) and MAGIC (red solid line and bow tie) are shown for TeV J2032+4130.}
\label{fig1}
\end{figure}

CTA may also shed light on the nature of VER J2019+407, which occupies a region of the northwest rim of G78.2+2.1 bright in radio and diffuse non-thermal X-rays \cite{2013ApJ...770...93A} that is less than $\sim 16\%$ of the area of the SNR.
As figure \ref{fig1} shows, the sources share a common power-law spectral index ($\Gamma \sim 2.4$) but differ widely in flux as well as angular extent.
These two sources may reflect accelerated particle populations with different energy cutoffs in different parts of the remnant, with the bulk of the remnant having a much lower cutoff (e.g. the green curve in figure \ref{fig1}) or VER J2019+407 may be something else entirely, such as a PWN in the line of sight.  CTA's better angular resolution and large field of view will make it much easier to disentangle the contributions from the remnant and VER J2019+407 and assess both their mutual relationship and their effect on the measured spectrum of the cocoon.

\begin{figure}[htbp!]
    \begin{subfigure}[t]{0.47\textwidth}
        \includegraphics[width=\textwidth]{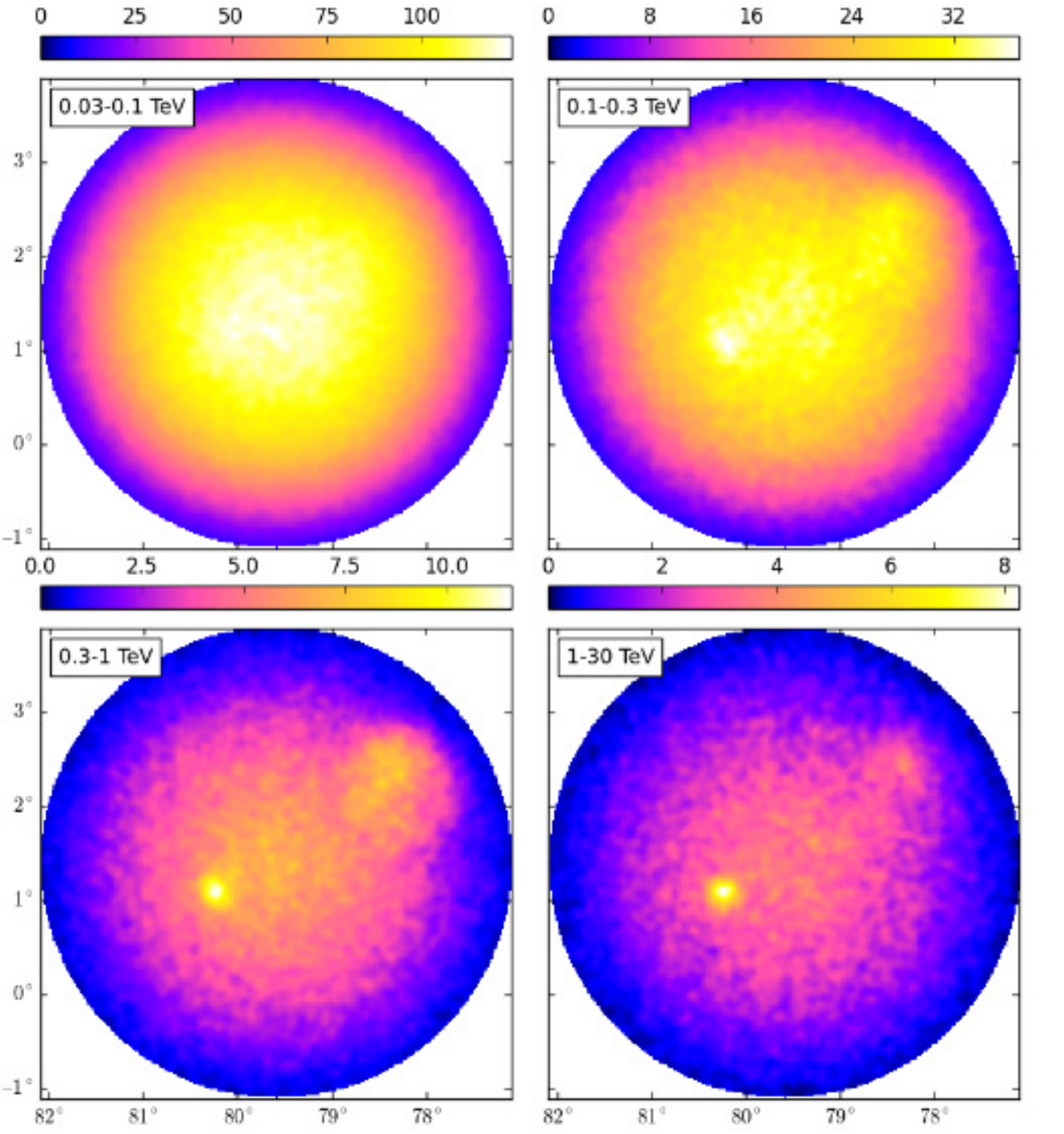}
        \phantomcaption{}
        \label{fig2}
    \end{subfigure}
   \begin{subfigure}[t]{0.47\textwidth}
        \includegraphics[width=\textwidth]{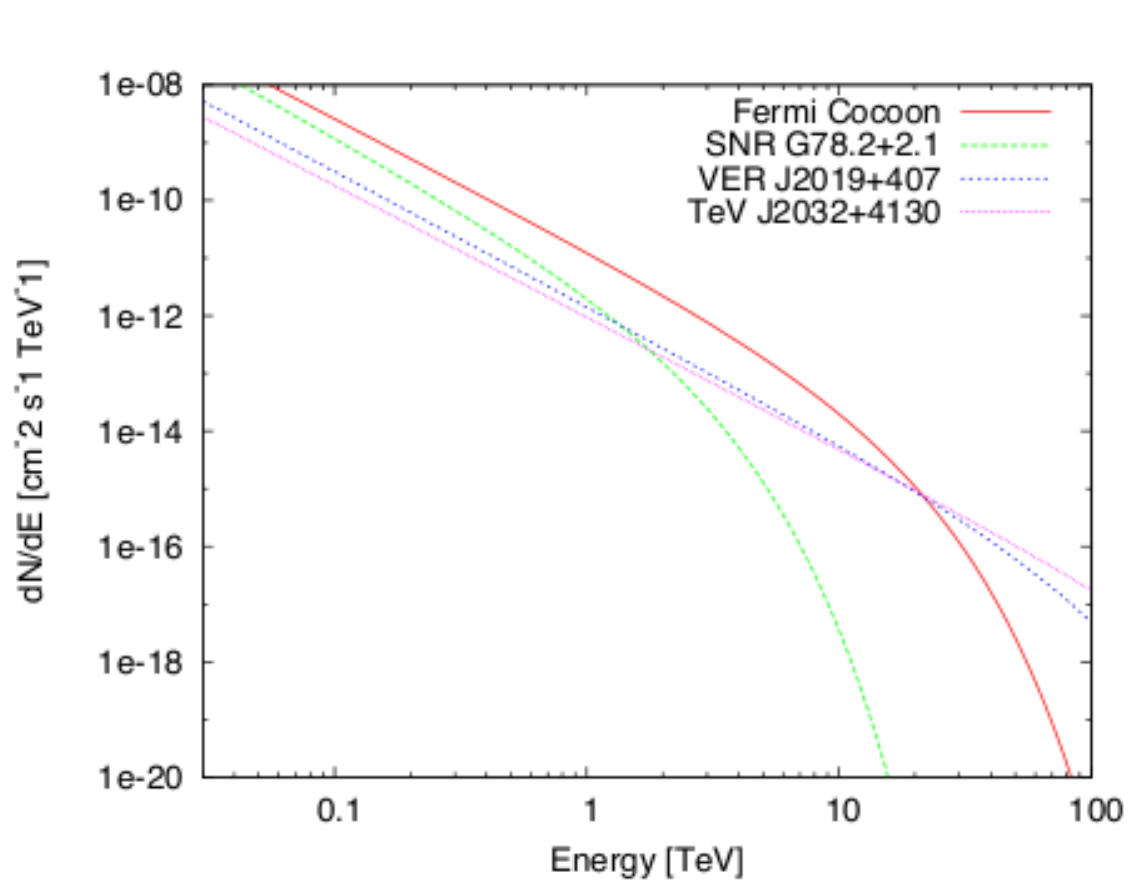}
        \phantomcaption{}
        \label{fig3}
    \end{subfigure}
\caption{\small Results from simulated deep-field observations of the Cygnus OB2/\F~cocoon region. Left: Simulated counts maps. Right:  Reconstructed spectra of known sources.}
\end{figure}

This is borne out by a dedicated simulation of a 100 hour observation of the Cyg OB2/cocoon region, processed through a binned likelihood analysis.
These simulations were run at the
Cyfronet computing facility of the University of Cracow using ctools for CTA
and the instrument response function was based on a site at Tenerife.
All known sources of VHE \g-ray emission above 10 GeV (the Fermi Cocoon, SNR G78.2+2.1, VER J2019+4007, TeV J2031+4130) were included and
all four  are detected with high ($>18\sigma$) significance, with a number of detected \g-ray counts which
exceeds 5\% of the residual CR background counts integrated over
the source area.
The count distribution as a function of energy (Figure \ref{fig2}) can be reliably used to determine arcmin-level positions and extensions of the
sources with few~$\%$ statistical uncertainties.
The spectra of the sources are also reliably reconstructed at the few~$\%$ statistical level, including all cutoff energies below 10 TeV (Figure \ref{fig3}). Precisely constraining cutoff energies above 10 TeV will require deeper exposures.

This work may benefit in future from a more advanced likelihood technique, which uses a \g/hadron classification parameter to gain additional separation power between photons and cosmic rays.
Such a technique improves overall sensitivity and compensates to some degree for influence of the cosmic ray background, which strongly impacts sensitivity below 300 GeV and contributes to the tendency of the fit to underestimate the extension of the more compact sources.
Nonetheless, we may be confident that the CTA observations of Cygnus will shed light on whether the
cocoon is a single source or the superposition of multiple objects and clarify
the relationship between VER 2019+407 and SNR G78.2+2.1.
The excellent spectral characterization will help establish whether the \g-ray emission is due primarily to hadronic or leptonic processes for these objects.
The measured cutoff energies will play a critical role in establishing maximum particle acceleration energies in the former case and probing energy losses in extreme environments in the latter.

\subsection{Cygnus OB1 and MGRO J2019+37}

The Cygnus OB1 region simulation considered known VHE $\gamma$-ray emitters MGRO J2019+37, VER~J2019+368, and VER~J2016+371.
In all cases the simulations extrapolated from the measured spatial and spectral properties of these sources.
The observed spectrum of the MGRO~J2019+37 seems to be best fitted with power law with exponential cutoff model.
The spectral index is found to be $\Gamma=2.0$ and the cutoff energy $E_c=29^{+50}_{16}$ TeV, with large error bars. No cutoff indication is given for VER J2016+371 and VER J2019+368.
In order to understand whether CTA will be able to study the morphology of the Cyg OB1 region and to resolve high energy cutoffs in the energy spectra of these sources, MGRO~J2019+37, VER~J2019+368 and VER~J2016+371 were each simulated with cutoff energies equal to $10$, $30$ and $80$ TeV.

\begin{figure}[htbp!]
    \begin{subfigure}[t]{0.5\textwidth}
        \includegraphics[width=\textwidth]{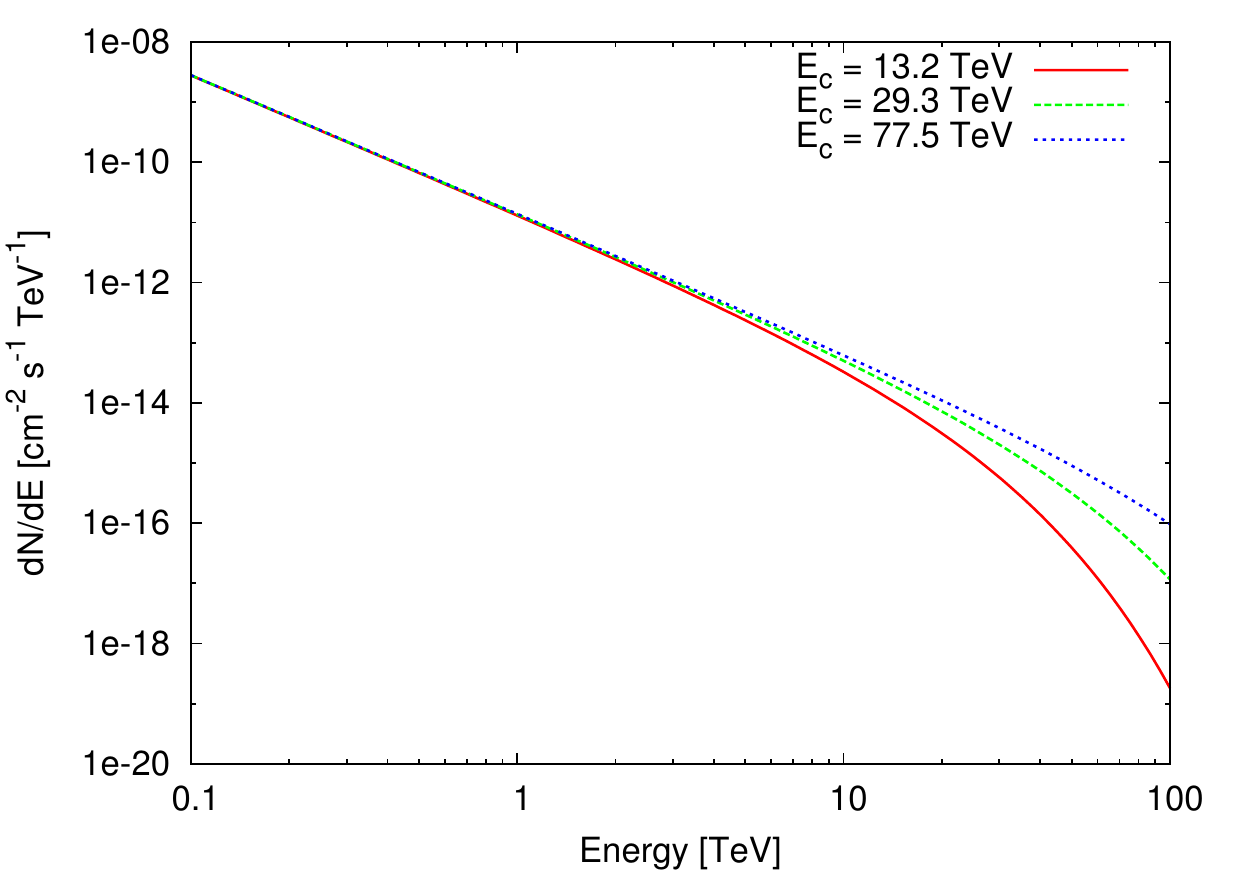}
        \phantomcaption{}
        \label{MGROJ2019}
    \end{subfigure}
   \begin{subfigure}[t]{0.5\textwidth}
        \includegraphics[width=\textwidth]{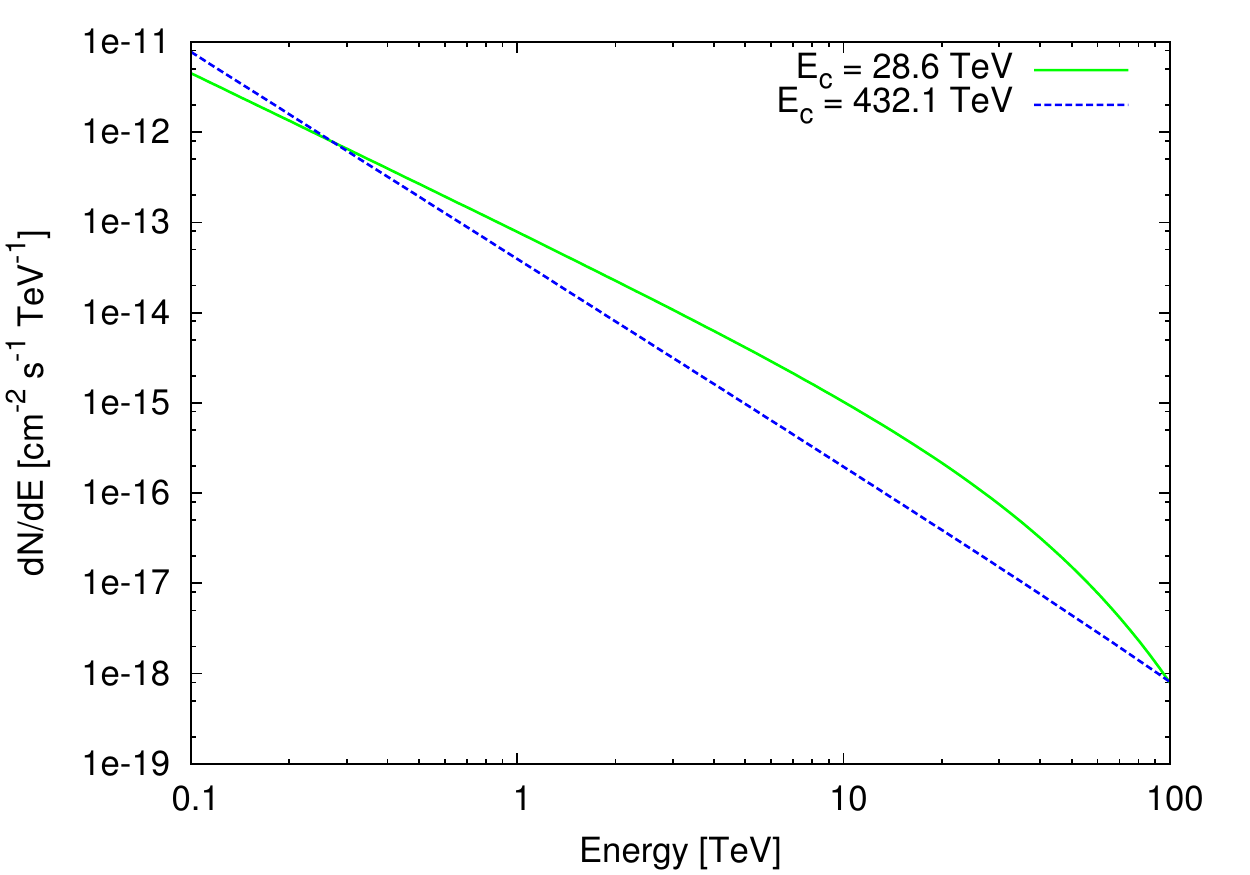}
        \phantomcaption{}
        \label{VERJ2019}
    \end{subfigure}
   \begin{subfigure}[t]{0.5\textwidth}
        \includegraphics[width=\textwidth]{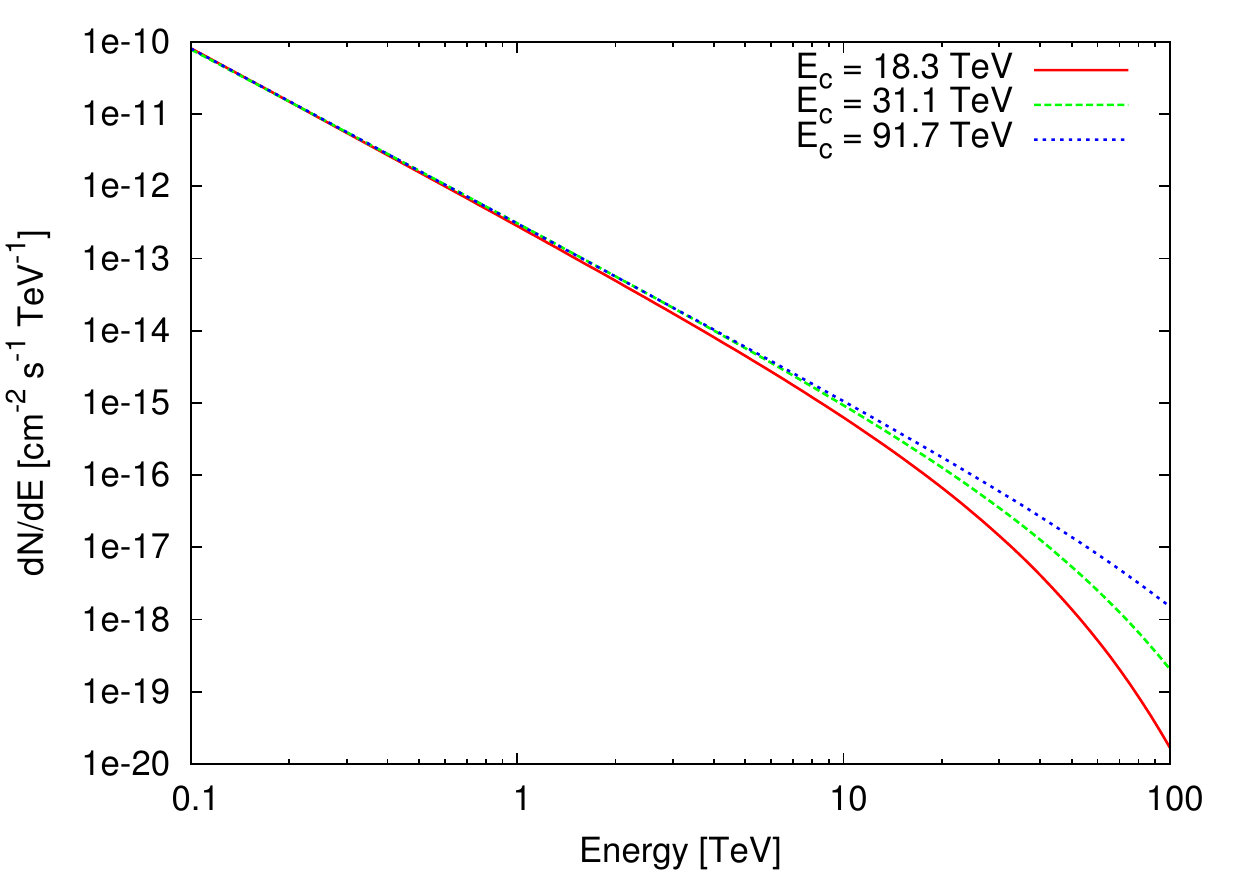}
        \phantomcaption{}
        \label{VERJ2016}
    \end{subfigure}
\caption{\small Spectra obtained from fits to simulated data for  MGRO~J2019+37 (top left), VER~J2019+368 (top right) and VER~J2016+371 (bottom) in the Cygnus OB1 region, each with three different assumed energy cutoffs (red, 10 TeV; green, 30 TeV; blue, 80 TeV).}
\label{fig:cutoffs}
\end{figure}

Spectra are extracted from the resulting simulated data using the same analysis as in Sec. \ref{sec:cygob2}; the spectra are shown in figure \ref{fig:cutoffs}.
The simulated cutoff energies for MGRO J2019+37 are reasonably well-reproduced when fit by power law model with an exponential cutoff: $13.2\pm 0.6$ TeV, $29.3\pm 1.9$ TeV and $77.5\pm 9.5$ TeV.
We anticipate that CTA has great potential to nail down the spectral cutoff of this source at high energies, where the VHE $\gamma$-ray emission is almost free of cosmic ray background.
The spectra of VER~J2019+368 and VER~J2016+371, however, fit better to a power law model.
Notably, the spectral analysis using an exponential cutoff failed for VER~J2019+368 when the assumed cutoff energy was chosen at 10 TeV, while the fitting errors exceeded 100\% for the 30 and 80 TeV cases.

\subsection{Physics of binary systems}

\textbf{Cyg X-3}, identified as the first GeV \g-ray-emitting microquasar, has opened up a new window on the non-thermal
processes associated with relativistic jets from accreting black holes.
The \F\/ orbital light curve of Cyg~X-3
suggests that GeV emission may originate from a region well-separated
from the binary system. It could be either due to the jets (via internal shocks)
or from a re-collimation shock when the jets interact with the dense wind from
the Wolf-Rayet
stellar companion.
At this stage, the \g-ray
spectrum of Cyg X-3 is still unconstrained at VHE.

Extrapolating the spectrum of the Cyg X-3 flare seen by \F{} \cite{2009Sci...326.1512F} at energies $> 30$ GeV we
estimate that a significant detection will take 200 min of observation time with 10 hours of exposure needed
for the determination of the full spectral shape.
It is therefore plausible that CTA's planned observations of Cygnus could trap a flare from either Cyg X-1 or Cygnus
X-3.
Persistent emission is reported with Fermi in the
case of Cyg X-3  \cite{2013ApJ...775...98B} with a flux of $\sim 0.5\times 10^{-6}$~ph cm$^{-2}$ s$^{-1}$ in the 0.1-10 GeV
range.
Without further spectral information extrapolation into the energy range probed by CTA is very difficult.
However, the few mCrab sensitivity promised by the CTA KSPs at the position of Cyg X-3 should strongly constrain both the level and nature of the high-energy \g-ray emission in the event of a non-detection.
CTA should be able to constrain the maximum energy reachable by this cosmic accelerator \cite{2013APh....43...81B} and potentially
localize the origin of the VHE emission.

\textbf{Cyg X-1:} INTEGRAL's recent detection \cite{2011Sci...332..438L} of strongly
polarized hard X-ray emission from Cyg X-1 supports the existence of high-energy
particles in this source.
A CTA detection of Cyg X-1 would help to bridge the gap
with stellar mass black holes in low mass X-ray binaries, which undergo regular outburst with a specific
phase of transient jet emission.
If we assume a Cyg X-1 flare with similar properties to that reported MAGIC ($dN/(dA dt dE)= 2.3 \time 10^{-12} (E/1
TeV)^{-3.2}$) flare \cite{2007ApJ...665L..51A}, we find from back-of-the-envelope estimates that a 5-$\sigma$ detection above 30 GeV can be achieved in 2
minutes. 10 minutes will be required to measure the spectral shape.
Persistent emission from Cyg X-1 has been detected
at 1 MeV  using INTEGRAL \cite{2011Sci...332..438L} but no corresponding detection has been reported by \F.
A naive extrapolation of the INTEGRAL detection to CTA energies predicts emission at the
mCrab level, roughly comparable to the sensitivity of the planned CTA GPS at this position.

\subsection{The Milagro Diffuse Emission}

We investigated the number intensity relation for HESS-detected VHE \g-ray sources above 20 mCrab in the region
of longitude $250^{\circ}<l<65^{\circ}$ and latitude $-3.5^{\circ}<b<3.5^{\circ}$ \cite{2013arXiv1307.4690C}. Extrapolating the number intensity curve down to 3-4 mCrab (the sensitivity planned for the Northern GPS) about 600-500 sources could be detected by CTA from
the HESS region. Rescaling this number for Cygnus as in
\cite{2008APh....29...63C}, faint sources resolved by CTA could account for about 40 per cent of the diffuse emission measured by
Milagro.
CTA can therefore better separate the true diffuse TeV \g-ray emission from unresolved sources and shed light on its nature.

\section{Acknowledgments}

We gratefully acknowledge support from the agencies and organizations
listed under Funding Agencies at this website: http://www.cta-observatory.org/.

\bibliographystyle{JHEP2}
\bibliography{Cygnus}

\end{document}